# DEEP LEARNING BASED MULTIPLE REGRESSION TO PREDICT TOTAL COLUMN WATER VAPOR (TCWV) FROM PHYSICAL PARAMETERS IN WEST AFRICA BY USING KERAS LIBRARY.

Daouda DIOUF[1], Awa Niang[1] and Sylvie Thiria[2]

[1]Laboratoire de Traitement de l'Information (LTI) – ESP – Université Cheikh Anta Diop de Dakar BP : 5085 Dakar-Fann (Sénégal)
[2]Laboratoire d'Océanographie et du Climat : Expérimentations et Approches Numériques (IPSL / LOCEAN) – Université Pierre et Marie Curie, 75252 Paris (France)

## ABSTRACT

Total column water vapor is an important factor for the weather and climate. This study apply deep learning based multiple regression to map the TCWV with elements that can improve spatiotemporal prediction. In this study, we predict the TCWV with the use of ERA5 that is the fifth generation ECMWF atmospheric reanalysis of the global climate. We use an appropriate deep learning based multiple regression algorithm using Keras library to improve nonlinear prediction between Total Column water vapor and predictors as Mean sea level pressure, Surface pressure, Sea surface temperature, 100 metre U wind component, 100 metre V wind component, 10 metre U wind component, 10 metre V wind component, 2 metre dew point temperature, 2 metre temperature.

The results obtained permit to build a predictor which modelling TCWV with a mean abs error (MAE) equal to 3.60 kg/m$^2$ and a coefficient of determination R$^2$ equal to 0.90.

## 1. INTRODUCTION

Water vapor is the most abundant greenhouse gas and is a good factor for the weather and climate [1][2]. The heating rate and circulation of the atmosphere depend greatly to the TCWV through the condensation of the latter into clouds. The atmospheric composition can be affected also by the TCWV via the photochemical reactions. A good prediction and monitoring of weather, climate and a better understanding atmospheric physics and chemistry go through a better knowledge of the TCWV that is highly variable in space and time.

At present, TCWV, also known as TPW (Total Precipitable Water), is retrieved from various imager remote sensing as AMSU on board the POES and METOP polar-orbiting satellites, SSM/I on board the DMSP F-13 satellite, from the sounders as GOES and ground-based Global Positioning System (GPS) equipment [3], [4].

The aim of this paper is to predict the Total Column water vapor(TCWV) from climate parameters in West Africa(*Figure 1*). The following variables were used: Mean sea level pressure, Surface pressure, Sea surface temperature, 100 metre U wind component, 100 metre V wind component, 10 metre U wind component, 10 metre V wind component, 2 metre dewpoint temperature, 2 metre temperature.

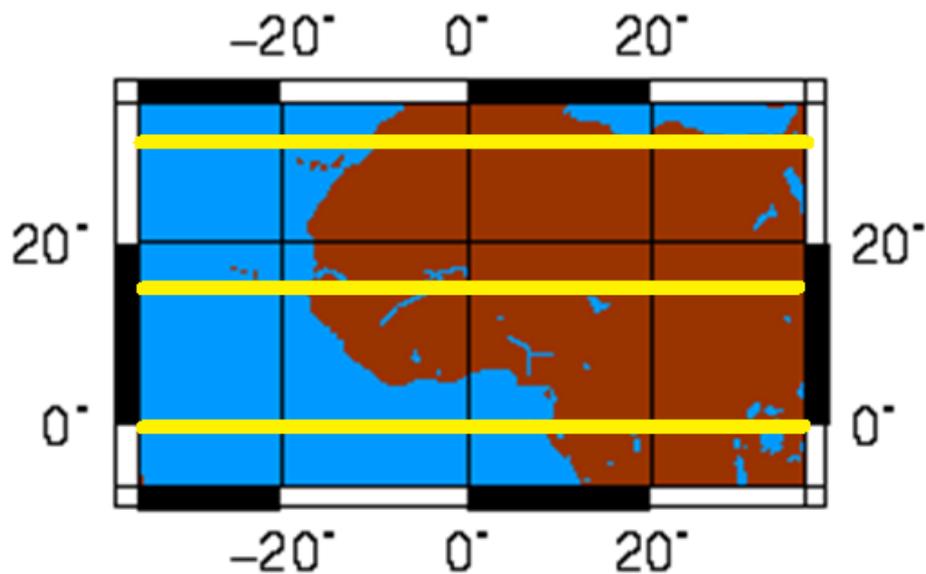

Figure1: Area of study and transect (in yellow) on latitude 0°N, 15°N and 30°N

Due to its high computing power, machine learning has shown a particular interest in processing and understanding of large and multifunctional data [5]. In the case of environmental data, these are often complex and highly non-linear. From this nonlinearity and complexity of data, we aim to build a deep learning model able to model the TCWV parameter from other parameters.

## 2. DATASET

The dataset we used are from European Centre for Medium-Range Weather Forecasts, ERA5 Reanalysis. These dataset are taken in an area of the West Africa, between -5°N and 34°N and -34°W and 35°W. These measurements extend the period of January 2004 to October 2018.
The learning dataset describes nine (09) and is concerned with modeling the Total column watervapor. These 09 parameters are noted by x and the TCWV by $y$.

To avoid the over fitting, we random the data.
Then 36 370 741 of pixel taken from 2004 to 2006 are random. From these random value, we take the 1% for train data and 0,5 % for test data.
The model is trained to predict the outputs and generalize to other non-trained data. Test data is used to test the accuracy of the model. It is to build, by learning, a neural model able to find the TCWV from input data.
For comparison we also used monthly observations data on GOME-2 instrument on board of the MetOp-A satellite.

## 3. NEURAL NETWORK MODEL

A deep learning based multiple regression network that consist an input layer, a multi-hidden layer with more the one hidden layer and an output layer. The nodes are fully connected. The number of input layer nodes is equal to the number of features of the input data. The more hidden layers, the higher the number of features needed to reduce the influence of under fitting or over fitting. Each hidden layer node is composed of neurons. The neurons contain both rectifier activation and aggregation function, when constructing the deep learning multiple regression model, the activation function in the default neuron is the Rectified Linear activation function,

14

making the deep learning network neurons have sparse characteristics, which reduces the influence of overfitting while increasing the depth of the network, improving the training speed of the model, and effectively overcoming the problem of gradient disappearance. This function that we must define is responsible for creating the neural network model to be evaluated [6].

## 3.1 Deep Learning Based Regression

A deep learning estimator is essentially based on the distributed representation, this mean that an output data is due to the interactions of various component sat different levels [7]. In this study, the deep learning estimator is organized in two training procedures, with a pre-learning and tuning with respect to the target TCWV.

## 3.2 Neural network model

We train the neural network by defining a sequential keras model. We are using the 09 inputs variables as Mean sea level pressure, Surface pressure, Sea surface temperature, 100 metre U wind component, 100 metre V wind component, 10 metre U wind component, 10 metre V wind component, 2 metre dewpoint temperature, 2 metre temperature. These 09 input features are fully connected to a first dense hidden layer of 64 (L1), this one fully connected to a second hidden layer of 32 neurons (L2), and finally using the activation function, the Rectified Linear Unit (ReLU), to process the output (Total column water vapor). ReLU are defined as f(x) = max(0,x) and are used with minibatch size of 64.

The workflow for training the model is simple.

We want to estimate a $y = g(x)$ function $(x \in R^p \quad et \quad y \in R)$ but by knowing only certain realizations of this function: $\{x_n, y_n\}$ $(n \in \{1...N\})$. This set is called learning set. The purpose of the learning is to estimate the weights of the network so that the output function noted $F$ best approaches the realizations of $g$. It is therefore a question of minimizing the following function so-called cost function:

$$J(w) = \sum_n \left\| y_n - F(x_n, w) \right\|^2 \text{ where } w \text{ is the set of weights.}$$

Since the cost function is the sum over all the realizations $\{x_n, y_n\}$, the gradient must be calculated for each of the realizations. Note $J^n$ the partial cost function corresponding to the realization n: $J^n = \left\| y_n - F(x_n) \right\|^2$

Let the error observed $J^n$ for the output neuron j and the training data n. The gradient with respect to the output $y_j$ of the neuron is:

$$\delta_j = \frac{\partial J^n}{\partial y_j}$$

Indeed, knowing the gradient with respect to the outputs of all the neurons of a layer k makes it possible to calculate the gradients with respect to the outputs of the neurons of the antecedent layer k-1:

$$\delta_i^{k-1} = \sum_j \frac{\partial J^n}{\partial y_j^k} \cdot \frac{\partial y_j^k}{\partial y_i^{k-1}} = \sum_j \partial_j^k \cdot w_{ij}^{k-1} \cdot f^{'}(v_j^k)$$

But it is easy to know the gradient of the cost function with respect to the output neuron. In our case, the quadratic cost function is:

$$\frac{\partial J^n}{\partial Y} = \frac{\partial (y_n - Y)^2}{\partial Y} = 2.(y_n - Y)$$

And so by backward propagation, first in the output layer, then in the hidden layers, we can calculate the gradient $J^n$ with respect to each of the weights of the network.

### 3.3 Tune the neural network

We have specified 140 epochs for our model. For this deep learning model, we choose Adam as an optimization algorithm [8]. Adam is an optimization algorithm that can used instead of the classical stochastic gradient [9] descent procedure to update network weights iterative based in training data. Adam is combining the advantages of two other extensions of stochastic gradient descent, specifically the Adaptive Gradient Algorithm (AdaGrad) and Root Mean Square Propagation (RMSProp).

## 4. RESULTS

Test data is used to test the prediction accuracy of the model. This model is used to predict TCWV from dependent or independent variables.

The accuracy on the learning set is 90.47% and the validation accuracy is 90.23%. The learning mean abs error is 3.60 kg/m² and the validation mean abs erroris 3.45 kg/m². In the figures below, the scatter plot between the target retrieved from training features and the real target are quite good. Most of the prediction error less than |5 kg/m²|.

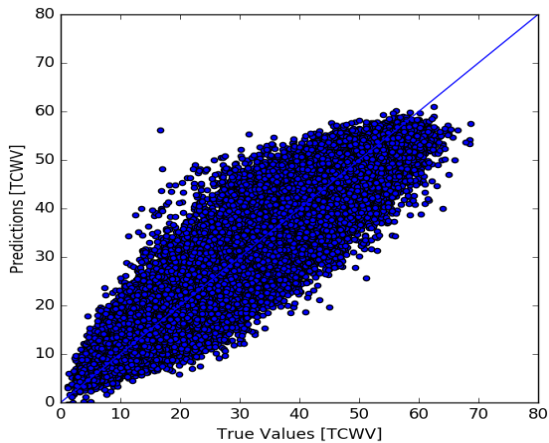

Figure 2: Scatter plot of predicted TCWV versus true TCWV

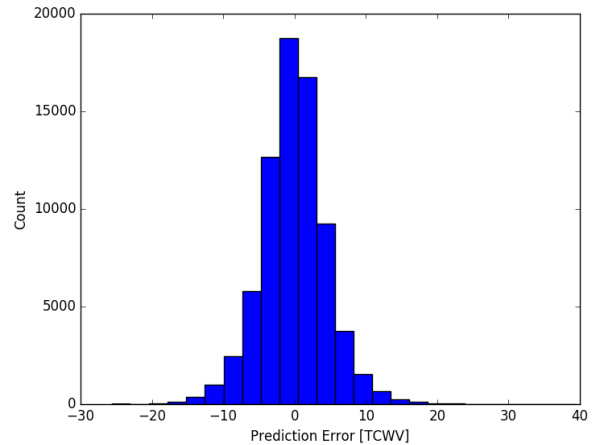

Figure 3: Error prediction

### 4.1 Validation against dependent data sets

We compared two data sets of total column water vapor that did not participate in learning phase to the measure from ERA-5 at the same date. Figure 4 show a comparison of TCWV predicted and TCWV measured above both land and ocean on January 2004. The global mean bias between the two data sets is quite small: 0.10 kg/m². Then, the TCWV retrieval from the others parameters by using neural network are obtained with good accuracy. January mean TCWV ranges from 0.5 to 57 kg/m². We denoted maximum values between -5°N to 5°N.

## 4.2 Validation against independent data sets

Comparison between the Total Column Water Vapor (TCWV) retrieved with the GOME-2 instrument on board of the MetOp-A satellite(c), the retrieved TCWV from model with using the ECMWF ERA-5 parameters reanalysis (b) and the measured TCWV of ECMWF ERA-5 (a) in May 2007 can be seen in figure 5. The patterns for the three boxes are very similar. We can observe that the highest values are all located between -5°N and 10°N.

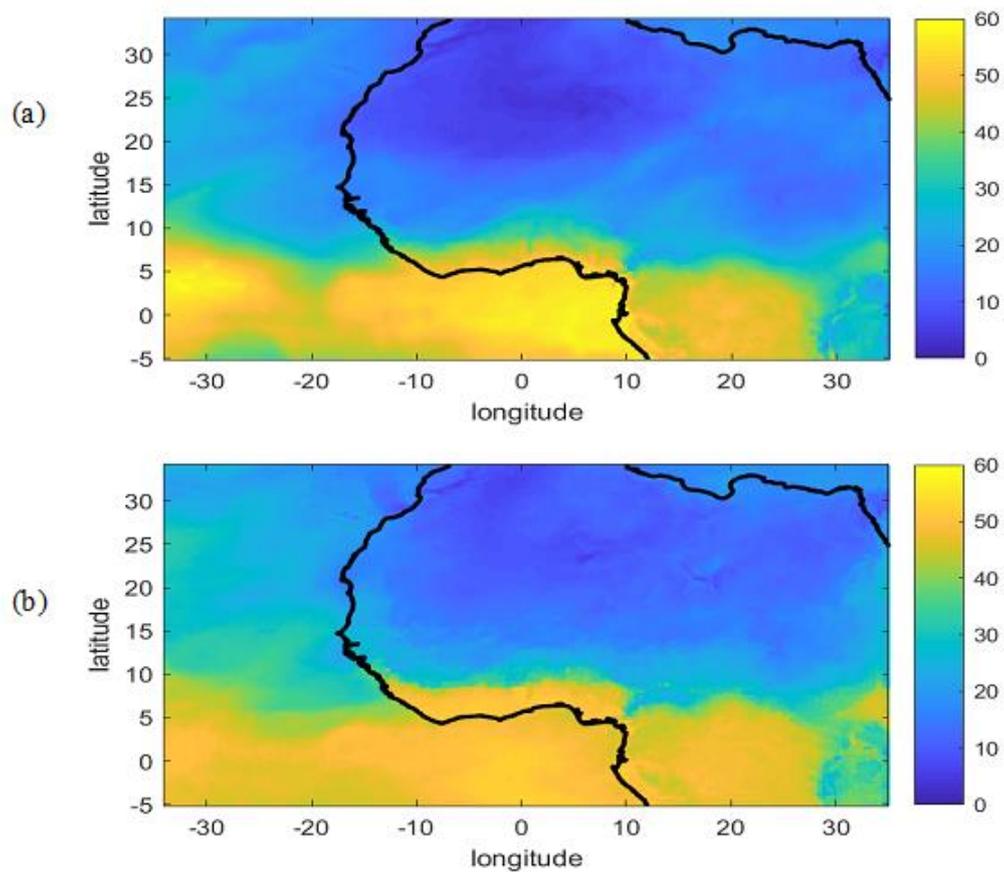

Figure 4: Map of the TCWV ECMWF ERA-5 analysis (a) with the corresponding retrieved from the neural network model(b) in January 2004.

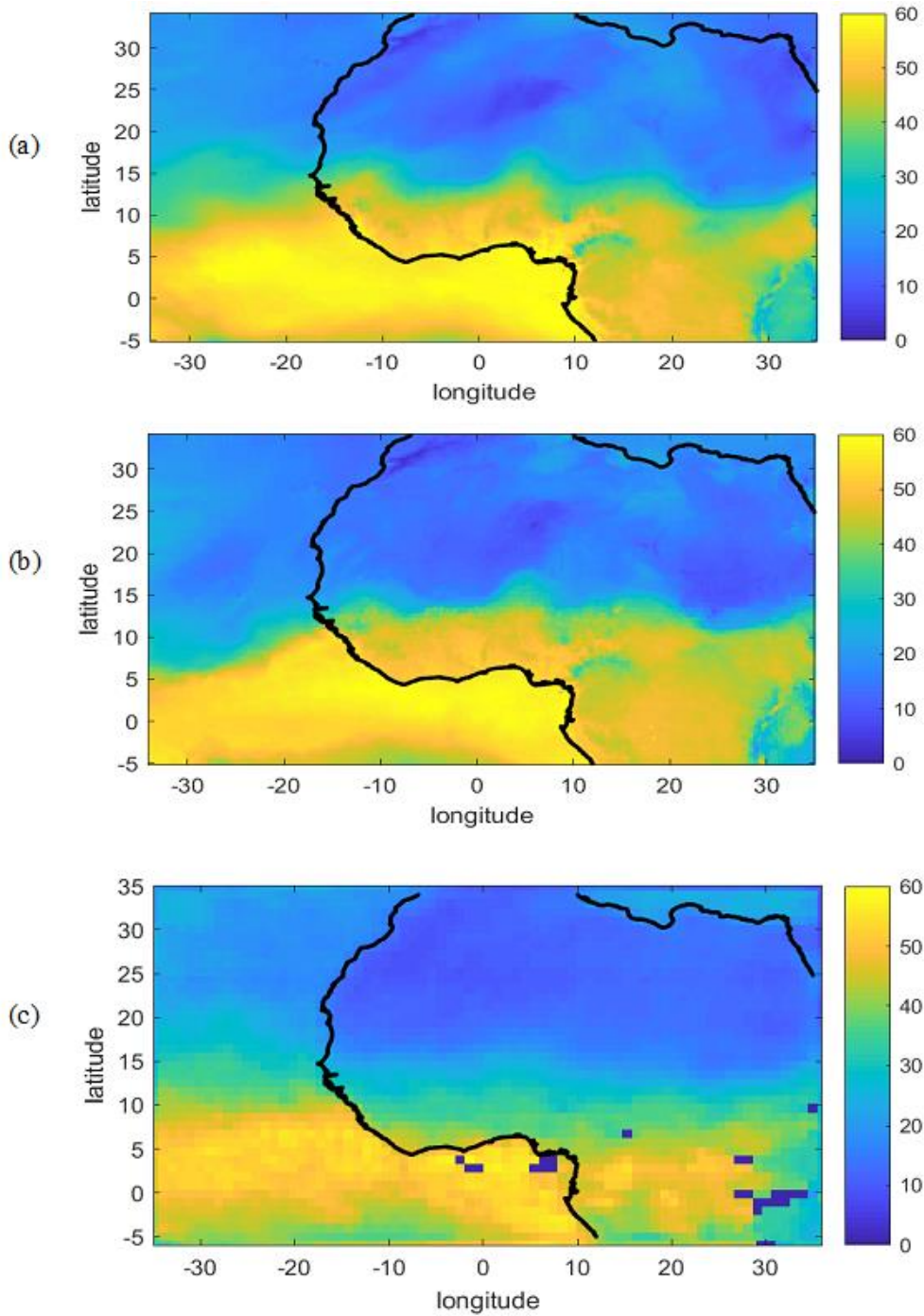

Figure 5:Map of the TCWV ECMWF ERA-5 analysis (a) with the corresponding retrieved from the neural network model (b) and GOME2 observations data (c) in May 2007.

The results shows the accuracy of the neural model to retrieved total column water vapor from few parameters. The figure 5 permit us to see that there is not more difference between the measured values (a) and the predicted values (b) but these last two have little difference with (c). We can see that the water vapor patterns over land and ocean are clearly visible with moist Intertropical Convergence Zone near the equatorial regions.

We are plotting the annual TCWV average retrieved for years 2004 and 2005. The plots concern the latitudinal transect at 0°N, 15°N and 30°NofTCWV, outputted by the neural network model using the annual average of the nine parameters as inputs for years 2004 and 2005, and compared by the annual TCWV ECMWF ERA-5 analysis average(figure 6). We also calculate the corresponding performances between the predicted annual TCWV average and the annual TCWV ECMWF ERA-5 analysis average at three latitudes for years 2004 and 2005 (Tab.1 and Tab.3). In addition, the performance of predicted annual TCWV average and those of the GOME2 observations TCWV are calculate and compared (Tab.2 and Tab.4). For these correlations, there are all quite high (> 90%) except at latitude 0°N when they are around 60-70%. From tab.1 to tab.4, we can observe the lowness of standard deviation for the years 2004 and 2005.

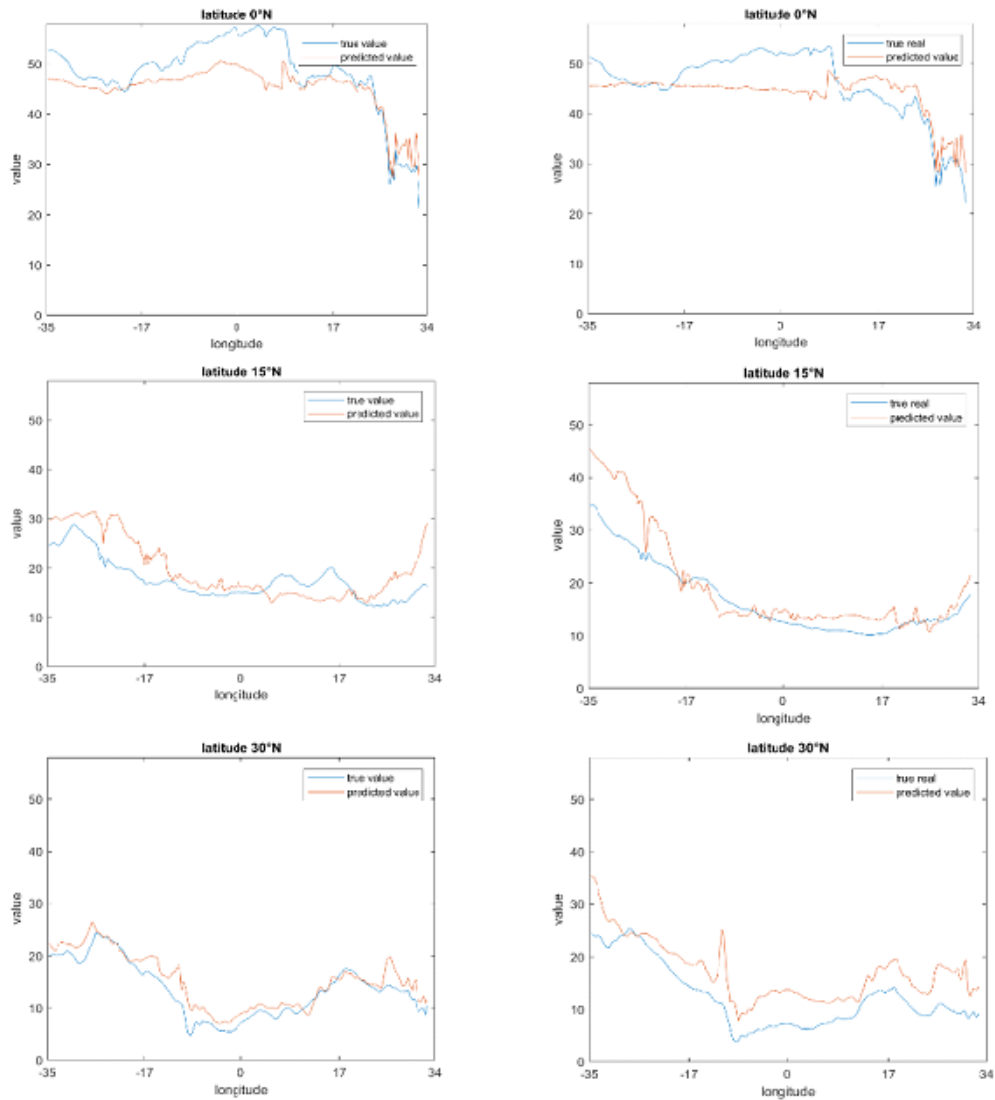

Figure 6: AnnualTCWV average for year 2004 (left) and 2005 (right) at different latitudes

Tab.1: Predicted TCWV vs.TCWV ECMWF ERA-5 analysis for 2004

|  | Standard deviation (kg/m2) | Correlation (%) |
|---|---|---|
| Latitude 0° N | 6.73 | 93.32 |
| Latitude 15° N | 4.52 | 76.68 |
| Latitude 30° N | 5.18 | 94.78 |

Tab.2: Predicted TCWV vs.TCWV GOME2 observations data for 2004

|  | Standard deviation (kg/m2) | Correlation (%) |
|---|---|---|
| Latitude 0° N | 2.61 | 60.33 |
| Latitude 15° N | 7.76 | 91.59 |
| Latitude 30° N | 5.12 | 90.44 |

Tab.3: Predicted TCWV vs.TCWV ECMWF ERA-5 analysis for 2005

|  | Standard deviation (kg/m2) | Correlation (%) |
|---|---|---|
| Latitude 0° N | 6.23 | 76.62 |
| Latitude 15° N | 7.75 | 95.26 |
| Latitude 30° N | 6.42 | 92.21 |

Tab.4: Predicted TCWV vs. TCWV GOME2 observations data for 2005

|  | Standard deviation (kg/m2) | Correlation (%) |
|---|---|---|
| Latitude 0° N | 2.61 | 63.4 |
| Latitude 15° N | 7.57 | 93.06 |
| Latitude 30° N | 4.6 | 92.05 |

## 5. CONCLUSION

In this paper, a focus was made on the ability of deep learning to predict the TCWV by using geophysical parameters as Mean sea level pressure, Surface pressure, Sea surface temperature, 100 metre U wind component, 100 metre V wind component, 10 metre U wind component, 10 metre V wind component, 2 metre dewpoint temperature, 2 metre temperature. We analyze the retrieved TCWV and compare its results with Gome2 observations. There are high precision with a mean global bias equal to 0.10 $km/m^2$ and the MAEis 3.41 $kg/m^2$. The annual prediction average of TCWV for three transects at 0°N, 15°N and 30°N compared to real measurement show good result about the effective of the deep neural regression model.

### Acknowledgements

European Centre for Medium-Range Weather Forecasts. 2017, updated monthly. ERA5 Reanalysis. Research Data Archive at the National Center for Atmospheric Research, Computational and Information Systems Laboratory. https://doi.org/10.5065/D6X34W69. Accessed 05 Feb. 2019.